\documentclass[10pt,conference]{IEEEtran}
\usepackage{amsmath}
\usepackage{amssymb}
\usepackage{graphicx}
\usepackage{subfigure}
\usepackage[linesnumbered,ruled]{algorithm2e}

\newcommand{\pluseq}{\mathrel{+}=}

\title{Denial of Service Attack Detection via Differential Analysis of Generalized Entropy Progressions}

\author{
\IEEEauthorblockN{Omer Subasi, Joseph Manzano, Kevin Barker}
\IEEEauthorblockA{Pacific Northwest National Laboratory (PNNL), Richland, WA, USA}
}

\begin{document}

\maketitle

\begin{abstract}
Denial-of-Service (DoS) attacks are one the most common and consequential cyber attacks in computer networks.
While existing research offers a plethora of detection methods, the issue of achieving scalability, a low false positive rate, and
high detection accuracy remains open.
In this work, we address this problem by developing a differential method based on generalized entropy progression.
In this method, named as DoDGE, we continuously fit the line of best fit to the entropy progression of destination addresses and check if the derivative, that is, the slope of this line
is less than the negative of the dynamically computed standard deviation of the derivatives. Furthermore, to distinguish from flash events, we leverage the symmetry that when a flash event occurs, the derivative of the entropy progression of source addresses is positive. 
With this design, we omit the usage of the thresholds and the results with five real-world network traffic datasets 
confirm that DoDGE outperforms threshold-based DoS attack detection by 
two orders of magnitude in terms of false positives on average. 
When compared to ten machine learning (ML) models, DoDGE achieves a balanced accuracy of 99\%,
while the average balanced accuracy for the ML models is 52\%.
Moreover, the results show that DoDGE successfully differentiates between a flash event and a DoS attack.
\end{abstract}

\section{Introduction}
As 5G wireless networks become mainstream, the cybersecurity of these 
networks becomes a first-class consideration \cite{10.1145/3278161.3278166}. Among many types of cyber attacks, 
Denial-of-Service (DoS) attacks \cite{doi:10.1177/1550147717741463}
are the most common and one of the most consequential types. 
The capabilities of 5G networks only make these attacks more difficult to defend against.
This challenge stems from the vast network bandwidth and density that 5G networks will provide and support.

There exists an abundance of research on the detection of DoS attacks \cite{doi:10.1177/1550147717741463} and \cite{Salim2019DistributedDO}.
There are many different approaches ranging from statistical approaches \cite{stat1} \cite{stat2} to machine and deep learning (ML and DL) \cite{mldl1, mldl2} to information-theoretical ones \cite{Basicevic}  \cite{BEHAL201796}  \cite{8778219}. 
Each type of these approaches has its own drawbacks. While ML and DL based methods perform relatively well in general, they are very expensive to train and 
require lots of data and very often finding representative data in sufficient quantity is very difficult.
This dearth of data in effect  hinders their scalability.  
Contrary to these methods, statistical and information-theoretical are relatively computationally cheap and scalable, 
however they require the explicit optimization of hyper-parameters
such as thresholds to signal an attack. In addition, usually their performance in terms of accuracy is less than that of learning based methods.
Moreover, they tend to have high false positive rates which incur prohibitive costs of disrupted services and lost capital only because legitimate traffic is falsely flagged as an attack. Therefore, the reduction of false positive rate should be treated as a major algorithmic goal. This is especially true since DoS attacks are rare considering the entire volume of Internet and mobile networks traffic.

In this work, we propose a \textbf{Do}S attack detection method using \textbf{D}ifferential analysis of \textbf{G}eneralized \textbf{E}ntropy progressions - \textbf{DoDGE}. It outperforms threshold-based methods by two orders of magnitude in terms of false positives
on average. 
In addition, DoDGE achieves a balanced accuracy of 99\%, whereas all ten ML/DL models tested have a balanced accuracy less than 62\%. The average balanced accuracy is 52\% for the ML models.
Moreover, contrary to learning based methods, DoDGE is lightweight, having both linear time complexity and a predictable memory footprint; and scalable, being embarrassingly parallel.
DoDGE achieves these properties while performing differential analysis of entropy progression.

DoDGE is built upon the postulation that when a DoS attack happens, the entropy of the destination addresses decreases. This postulate is justified since
DoS attacks by construction target specific destination addresses thereby decreasing the entropy of such addresses. This assumption is nearly always observed in real-world scenarios. On other hand, there exist benign events that decrease destination entropy in a similar manner. For example, flash events (crowds), such as a sporting or an entertainment event, occur where legitimate users access certain destinations, such as the organizing website, during the event in large numbers. This can easily and falsely be taken as a DoS attack by detection algorithms even though it is legitimate traffic. To distinguish from flash events, we leverage the symmetry that when a flash events occurs, the entropy progression of source addresses increases. This effectively means the derivative of the entropy progression of source addresses is positive as opposed to the derivative of destination progression being negative during a DoS attack. We observed this symmetry in the flash events of France 98 World Cup \cite{france98} and posit that it is a characteristic of flash events.

In summary, the main contributions of our study are:
\begin{itemize}
\item We developed an entropy based DoS attack detection algorithm utilizing differential analysis. DoDGE removes the threshold hyperparameter and leverages the derivative of entropy progression to improve its accuracy.
\item DoDGE leverages the existing (or lack thereof) symmetry between source and destination addresses to differentiate between attacks and flash events. 
\item DoDGE employs generalized entropies measures which are shown to be superior to Shannon entropy in terms of detection as shown in Section \ref{eval}.
\item Results show that, on average, DoDGE outperforms threshold-based methods by two orders of magnitude in five real-world network traffic datasets in terms of false positives. Moreover, DoDGE successfully locates flash events in five datasets taken from France 98 World Cup \cite{france98} with low false positive rates.
Compared to ten ML models whose highest balanced accuracy is 62\%, DoDGE achieves 99\% balanced accuracy.
\item DoDGE uses only local data and as a result it can be deployed in the 5G edge nodes and Internet routers. In effect, this placement makes it embarrassingly parallel. 
\end{itemize}

This paper is organized as follows: 
Section \ref{background} provides background on the concept of entropy and its generalized measures.
Section \ref{dosdetection} presents our novel DoS attack detection algorithm.
Section \ref{eval} discusses the experimental results.
Section \ref{related} overviews the related work.
Finally, Section \ref{conclusion} presents concluding remarks.

\section{Background}
\label{background}
\subsection{The Concept of Entropy}
The notion of entropy appears in many areas such as thermodynamics, information theory and statistical mechanics.
It generally refers to a measure of disorder, randomness, and uncertainty. 
In information theory, the most well-known entropy is Shannon entropy \cite{Shannon-ent}: $H(X) = - \sum_i^n  p_i \ log(p_i)$
where $X$ is a discrete random variable which has possible outcomes $x_i$ with probability $p_i$.

Complex dynamical systems having multifractality, systems with long range forces, 
and entanglement in quantum systems require generalized entropy measures with weaker assumptions than Shannon's entropy such as nonadditivity \cite{genentreview}.
Multifractal systems are those whose dynamics can be described by its subsystems such as turbulence, human brain activity, and geophysics.
Renyi entropy is one of the most well-known generalized entropy measures \cite{renyi1961measures}.
The Renyi entropy of order $\alpha$, $\alpha \geq 0$ and $\alpha \neq 1$ is defined as:
\begin{equation}
    \mathrm {H}_{\alpha}(X)={\frac {1}{1-\alpha }}\log {\Bigg (}\sum _{i=1}^{n}p_{i}^{\alpha }{\Bigg )}. \nonumber
\end{equation}
We note that as $\alpha \rightarrow 1$, Renyi entropy reduces to Shannon entropy.

Another important generalized entropy is Tsallis entropy \cite{tsallis1988possible}, 
which we use as the entropy measure in the proposed method.
The Tsallis entropy is defined as:
\begin{equation}
    S_{q}(X)=\frac{1}{1-q} {\Bigg (1-\sum_{i}^n p_{i}^{q}\Bigg )} \nonumber
\end{equation}
where $q$, $q \neq  1$, is a real parameter.
Similar to Renyi entropy, as $q \rightarrow 1$, Tsallis entropy converges to Shannon entropy.

In this work, we leverage the parameter $q$ as a tool to magnify and highlight the entropies of network addresses.
Specifically, it can be used to highlight the entropy of (a) network addresses with no or low activity for $q < 1$, (b) a large portion of network addresses for $q$ in the vicinity of $1$, and (c) network addresses with high activity for $q > 1$. The last case explains why we set and fix the Tsallis parameter value $q=8$ (bigger than 1) to prevent false positives and we do not vary it in our experiments.
\subsection{Thread Model and Deployment of DoDGE}
\begin{figure}[ht!]
\centering
\includegraphics[scale=0.35]{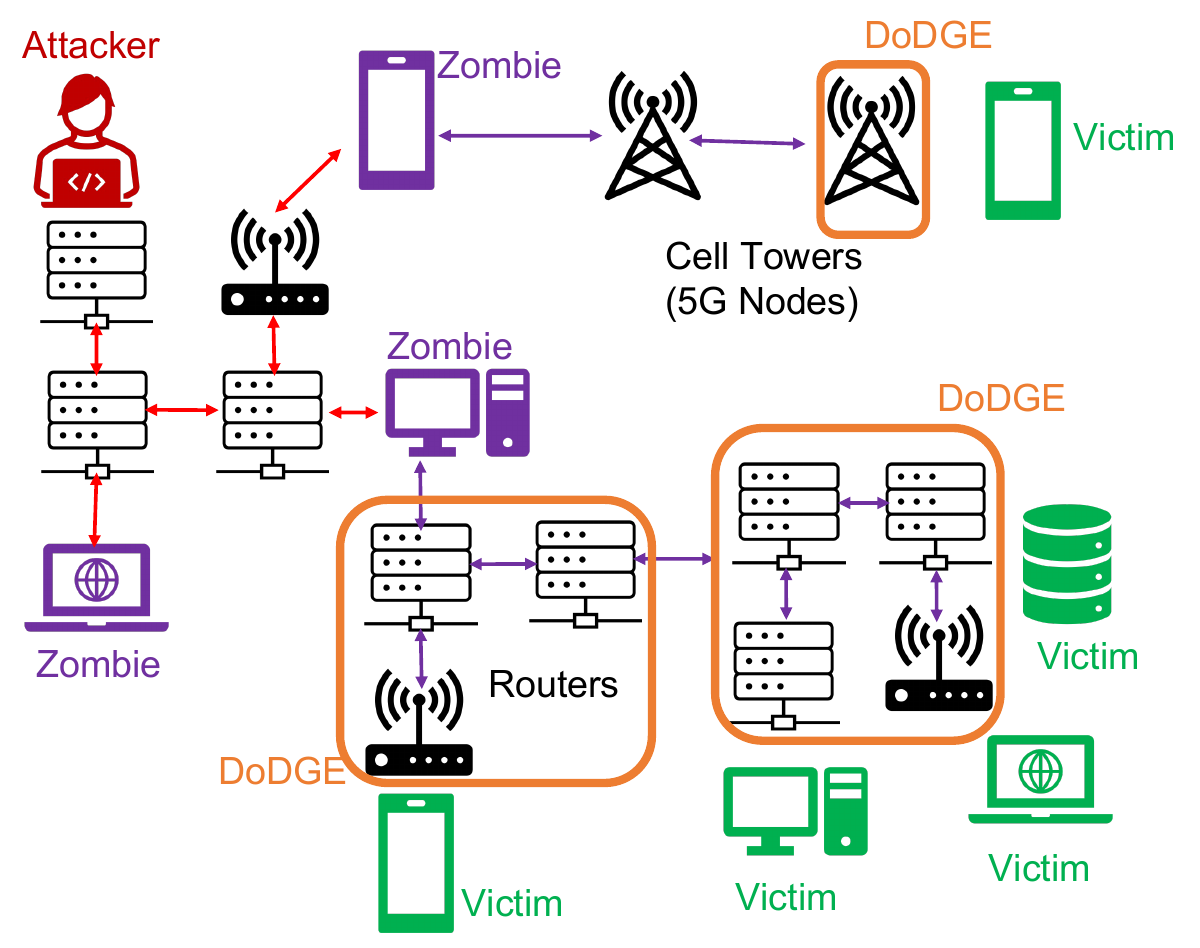}
\caption{Thread model for our study.}
\label{fig:threadmodel}
\centering
\includegraphics[scale=0.45]{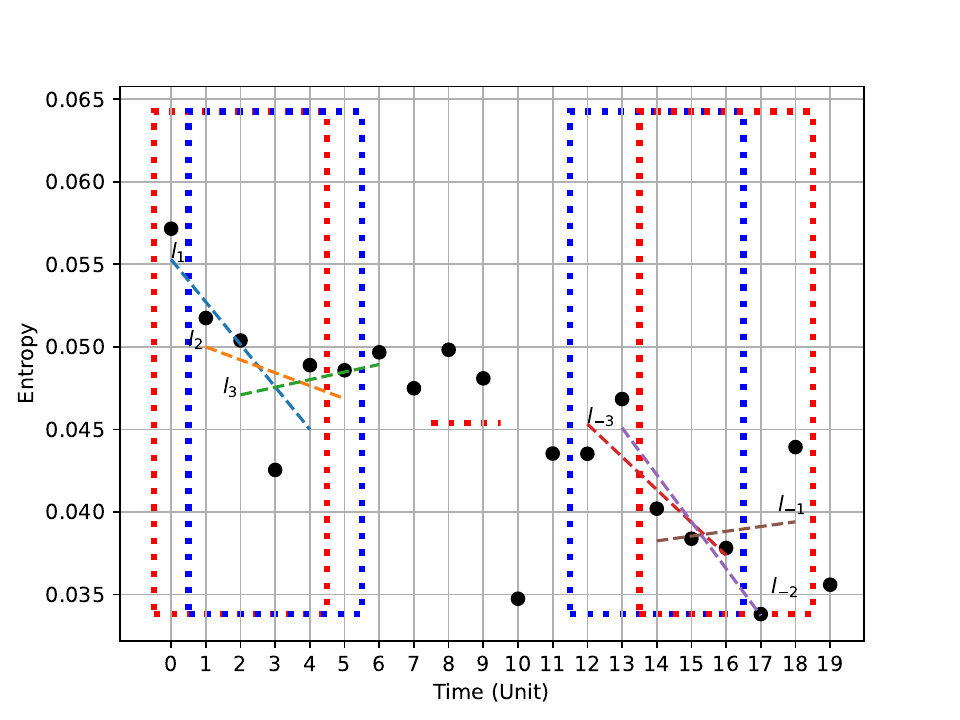}
\caption{An entropy progression with the successive lines of best fit computed and drawn.}
\label{fig:pnderivative1}
\end{figure}

Figure \ref{fig:threadmodel} illustrates our thread model. Here, an attacker exploits other computing systems connected to the Internet and launches a DoS attack.
DoDGE is placed at 5G nodes or cell towers and Internet routers. When deployed at the 5G nodes, DoDGE operates completely local, i.e, as a single non-communicating process. When deployed at routers, DoDGE exchanges messages to a small group of neighbors (3-4). An attack is signaled when majority of them detect one. Each DoDGE instance runs locally but to reach a common decision a small number of messages are exchanged. 

\section{The DoDGE Algorithm}
\label{dosdetection}
In this section, we discuss the design of DoDGE.
Then, we discuss the different computational tricks that we use to reduce DoDGE computational costs.
Finally, we study the computational and memory complexity of our method.
\subsection{Design of DoDGE}
\label{design}
The DoDGE algorithm is based on calculating the derivative of the progression of recent entropies. 
The entropy progression is the time series of the entropies computed based on destination or source IPs. 
The progression is a small number of past entropies. 
 For example, Figure \ref{fig:pnderivative1} shows an entropy progression of size 5 which is synthetically generated with Tsallis entropy.

We use the destination addresses' entropy to detect DoS attacks.
To detect an entropy decrease, we check if the derivative of the entropy progression is negative.
To calculate this derivative, we use the simplest model: line of best fit. 
The slope of this line is the derivative of the progression. If the derivative is negative, then the entropy is decreasing.

We not only look for a negative derivative but an \emph{outstandingly negative} one. For this purpose,
we compute dynamically the standard deviation of the derivatives to increase the precision of detection.
An attack is signaled when the derivative of the progression is less than the negative of the standard deviation.

In our design, in addition to Shannon entropy \cite{Shannon-ent}, we employ and experiment with two generalized entropies: 
Renyi \cite{renyi1961measures} and Tsallis \cite{tsallis1988possible} entropies. 
We use generalized entropies to amplify the magnitude of the computed entropy. 
This improves the method's precision and accuracy. In Section \ref{eval}, we include results that demonstrate
this effect. 

Moreover, DoDGE does not use any thresholds to signal an attack.
This is because of the inherent limitations of threshold-based attack detection methods - especially those using static or rigid thresholds. 
They often produce sub-optimal results (as seen in Section \ref{eval}) which DoDGE alleviates. 

We design DoDGE to be able to distinguish between DoS attacks and flash events 
by utilizing their symmetrical behavioral characteristics. 
We develop DoDGE based on the following observations of real-world flash events traffic: 
When a flash event occurs, the entropy of source addresses increases. This basically corresponds to a positive derivative. Moreover, we observe that during flash events the entropy of source addresses are bigger than that of destination addresses. This makes sense because during flash events typically the number of different users, that is source addresses, increases, which results in an increased source entropy. One final observation is that during flash events the number of source addresses are most often bigger than the number of destination addresses. 
\SetAlgoNoEnd\SetAlgoNoLine
\DontPrintSemicolon
\begin{algorithm}[ht]
\caption{DoDGE Pseudocode}\label{alg:main}
\KwData{The Destination Progression $\{EP_{D_i}\}$}
\KwData{The Source Progression $\{EP_{S_i}\}$}
\KwData{Current Destinations Window $W_D$}
\KwData{Current Sources Window $W_S$}
\KwData{Entropy Function $H$}
\KwResult{\textbf{True} if a DoS attack is detected}
$mean\_prev = mean\_curr = cnt = var = 0$  \texttt{\\}
 \While{(True)}{
 $destination\_entropy = H(W_D)$ \label{nent}  \texttt{\\}
 $source\_entropy = H(W_S)$ \label{snent}  \texttt{\\}
 $\{EP_{D_i}\}.pop()$  \label{up1} \texttt{\\}
 $\{EP_{D_i}\}.push(destination\_entropy)$  \texttt{\\}
 $\{EP_{S_i}\}.pop()$  \label{sup1} \texttt{\\}
 $\{EP_{S_i}\}.push(source\_entropy)$  \label{up2} \texttt{\\}
 $m_d, b = \mathbf{line\_of\_best\_fit}(\{EP_{D_i}\})$ \label{line} \texttt{\\}
 $m_s, b = \mathbf{line\_of\_best\_fit}(\{EP_{S_i}\})$ \label{sline} \texttt{\\}
 $mean\_prev = mean\_curr$  \label{w1} \texttt{\\} 
 $mean\_curr \pluseq (m_d-mean\_curr)/cnt $ \texttt{\\}
 $var \pluseq (m_d - mean\_curr) * (m_d - mean\_prev)$  \tcc{Welford's method}
 $\sigma = math.sqrt(var/cnt)$ \label{wf}  \texttt{\\}
$cnt = cnt + 1$ \texttt{\\}
$diff\_ent = source\_entropy - destination\_entropy$ \label{diffent}  \texttt{\\}
$diff\_num = |EP_{S_i}| - |EP_{D_i}|$ \label{diffnum} \texttt{\\}
  \uIf{$m_d < - \sigma$}{\label{checkd}
  \If{$m_s > 0 \ \&\  diff\_ent > 0$}{ \label{fi} 
    echo "Flash Event"; \textbf{continue} \texttt{\\}
    
  }
  \uElseIf{$m_s > 0 \ \&\  diff\_num > 0$}{
    echo "Flash Event"; \textbf{continue} \texttt{\\}
    
  }
  \uElseIf {$diff\_num > 0 \ \&\ diff\_ent > 0$}{
    echo "Flash Event"; \textbf{continue} \label{ff} \texttt{\\}   
  }
  \Else {
  \textbf{Error} "DoS Attack" \texttt{\\}
  \textbf{return True} \label{checkdf}
  }
  } 
 }
\end{algorithm}

Algorithm \ref{alg:main} shows the pseudo code for our DoS attack detection approach.
Given destination and source entropy progressions, their current windows, and an entropy function, the algorithm
computes the destination and source entropy of the current windows (Lines  \ref{nent} and \ref{snent}).
It updates both destination and source progression queues with the newly computed entropies (Lines \ref{up1} - \ref{up2}).
Then it computes the destination and source derivatives ($m_d$ and $m_s$ respectively) of the line of best fit to the corresponding progression (Lines \ref{line} and \ref{sline}). 
After that, it uses Welford's method \cite{Welford62noteon} to 
compute the standard deviation of the \emph{destination} derivatives (Lines \ref{w1} - \ref{wf}).
Additionally, the algorithm calculates the difference between source and destination entropies (Line \ref{diffent}), and the difference between the number of different source and destination addresses (Line \ref{diffnum}).
The algorithm first checks if the destination derivative is less than the negative standard deviation (Line \ref{checkd}).
If it is, then it checks if any two of the following are positive: The derivative of the source progression, the difference between the source entropy and destination entropy, and the difference between the number different source and destination addresses. If so, then a flash event is echoed and the algorithm continues its normal execution. If the destination derivative is positive but two of the three conditions does not hold together, then a DoS attack is detected.

We note that there will be a problem regarding the numerical stability of the computation of the standard deviation since DoDGE computes the standard deviation of the derivatives dynamically. 
We remove this numerical stability problem by using \textbf{Welford's method} \cite{Welford62noteon}.

DoDGE samples the network flow randomly and performs analysis
on the sample.
Sampling reduces the computational and memory costs significantly.
We set the sampling ratio to 5\% in our design. 
Because the entropy calculation is a computational complex process, we reduce its costs by transforming the floating point operations to integer ones while using the appropriate scaling to preserve the relative magnitudes. Moreover, we omit unnecessary computations, such as the square root of the variance, since the exact results are not needed and the relative magnitudes are sufficient. 

\subsection{Complexity Analysis}
\label{complexity}
The computational complexity of DoDGE is linear in the size of network flow in the unit-time window. 
The computational cost of DoDGE results from the entropy computations, fitting the lines of best fit to the entropy progressions,
computing the standard deviation dynamically and checking the detection conditions. 
Let $N$ be size of network flow in the unit-time window.
The two (source and destination) entropy computations are $O(N)$.
Fitting the two lines of best fit to the entropy progression which has a fixed small number of entropies is $O(1)$.
Computing the standard deviation of the destination derivatives on-the-fly is $O(1)$.
Checking the detection conditions is $O(1)$. 
Finally, computing the frequencies of IP addresses via a map is $O(N)$.
Therefore, the total computational complexity is $O(N)$.
Sampling will reduce this linear complexity with a constant. In our case, 
it is about 20 which is the multiplicative reciprocal of 5\% sampling ratio.

Memory complexity is also linear in the size of network flow in the unit-time window. 
The memory for the unit-time window is $O(N)$, while the memory for a map is \emph{at most} $O(N)$.
The memory needed for the temporary variables is $O(1)$.
Therefore, the total memory complexity is $O(N)$.

\section{Evaluation}
\label{eval}
In this section, we present the experimental setup (Section \ref{setup}), the threshold-based detection framework (Section \ref{thresholdframe}), 
and finally main results (Section \ref{mainresults}).
\subsection{Experimental Setup}
\label{setup}
Our first five datasets are from \cite{JAZI201725}, \cite{ERHAN2020106187}, and \cite{Iman}.
These datasets are used particularly for DoS detection. They do not have flash events (crowds). 
We have an application layer dataset, which we name ``Application'', from \cite{JAZI201725}.
Twenty six DoS attacks were intermixed with the attack-free traces from \cite{SHIRAVI2012357} . 
This dataset contains 24 hours of network traffic with total size of 1.4 GB. 
``Application'' has application-layer DoS attacks that are both high and low volume. 
These attacks include high volume HTTP, Slow Send and Slow Read attacks.
We have two datasets from \cite{ERHAN2020106187} which are UDP and TCP traffic. We name them ``UDP'' and ``TCP'' datasets respectively.
TCP has a size of 955 MB while UDP has 795 MB.
Both datasets last about 8 minutes.
The other two datasets are from \cite{Iman}. They consist of benign and mixed traffic. We name them ``Benign'' and ``Mixed'' datasets respectively.
The Benign dataset includes only benign internet traffic for about 3 hours. Its size is 11 GB.
The Mixed dataset includes DoS attack traffic which lasts about 4 hours. It is 13 GB of mixed attack and benign traffic.

We use the labeled dataset from \cite{mlcompdataset} to compare DoDGE to ten different ML models. The dataset contains labeled benign traffic and the most up-to-date common DDoS attacks. We use the subset that has flood attacks. This subset is about 3GB in its CSV format.

Finally, we use five datasets from France World Cup 98 \cite{france98}. We use them to evaluate DoDGE's performance against flash events. 
They consist of all the requests made to the 1998 World Cup Web site (www.france98.com) between April 30, 1998 and July 26, 1998. The five datasets have a size of 20GB in total. 
These five datasets have only flash events 
- no DoS attacks. We evaluate DoDGE for detecting flash events with these datasets which have only benign traffic. Any DoS attack detection is a false positive as a result. In particular, we choose Days 48, 63, 66, 69, and 78. Although these datasets are old, the characteristics of flash events that they model, is not different than what we would have as a flash events nowadays. Many recent research studies use them \cite{8685891, BEHAL201796}.

In this study, we set the unit window (time) 1 minute for all datasets and the Renyi entropy parameter to zero.

\subsection{A Bidirectional Short and Long-term Entropy Method with Dynamic Thresholds}
\label{thresholdframe}
We evaluate DoDGE against a general method encapsulating various entropies, decision logic and dynamic thresholds.
Our aim is to explore the design space of entropy based methods that use dynamic thresholds.
Results in Section \ref{eval} confirm that no matter how general the method is, this method produces a very high number of false positives.

First, this method uses the notion of \emph{bidirectional entropy} which incorporates both the entropy of source and destination traffic flows.
Our assumption regarding the bidirectional entropy is the following: When there is a DoS attack,
the entropy calculated on the destination IPs and on the source IPs decreases.
Second, it maintains both short-term and long-term entropies. Short-term entropy is the entropy of recent windows, 
meanwhile long-term entropy is the entropy of windows that are further in the past.

Third, this general method maintains dynamic thresholds and accommodates different 
decision strategies (decision logic).
There are many possibilities for dynamic thresholds. For instance, $threshold_t = \sum_{j = t-k}^{t-1} threshold_j/k $
where $t$ is the current time and $k$ is some positive integer. Here, the threshold is the moving average of the last $k$ thresholds.
We use all past values of the threshold in the simulations.

A decision strategy is needed to decide when to signal an attack. It is used to raise an attack alert or not.
A decision strategy is defined as a Boolean-valued function whose input is a vector of entropies and thresholds. 
As an example,
\begin{align}
S &= dst_{ste} < dst_{stthr} \  \&  \ dst_{lte} < dst_{ltthr} \nonumber
\end{align}
where $dst_{ste}, dst_{lte}, dst_{stthr},$ and $dst_{ltthr}$ are destination addresses' short term and long term entropy, and 
short term and long term threshold respectively.
Similarly, $src_{ste},  src_{lte}, src_{stthr}$, and $src_{ltthr}$ source addresses'  
short term and long term entropy, and short term and long term threshold respectively.

We experiment with the following seven strategies:
\begin{align}
&S1 =  (dst_{ste} < dst_{stthr}) \  \& \  (dst_{lte} < dst_{ltthr}),  \nonumber \\
&S2 =  (src_{ste} < src_{stthr})  \  \&  \  (src_{lte} < src_{ltthr}), \nonumber \\
&S3 =  (dst_{ste} < dst_{stthr}) \  \&  \  (src_{lte} < src_{ltthr}),  \nonumber \\
&S4 = dst_{ste} < dst_{stthr}, \ S5 = src_{ste} < src_{stthr}  \nonumber \\
&S6 =  dst_{lte} < dst_{ltthr}, \ S7 =  src_{lte} < src_{ltthr} \nonumber
\end{align}
We implement a Python framework to evaluate all strategies and a C++ prototype for timing results.

\subsection{Results}
\label{mainresults}
\begin{figure*}[ht!]
\centering
\includegraphics[scale=0.33]{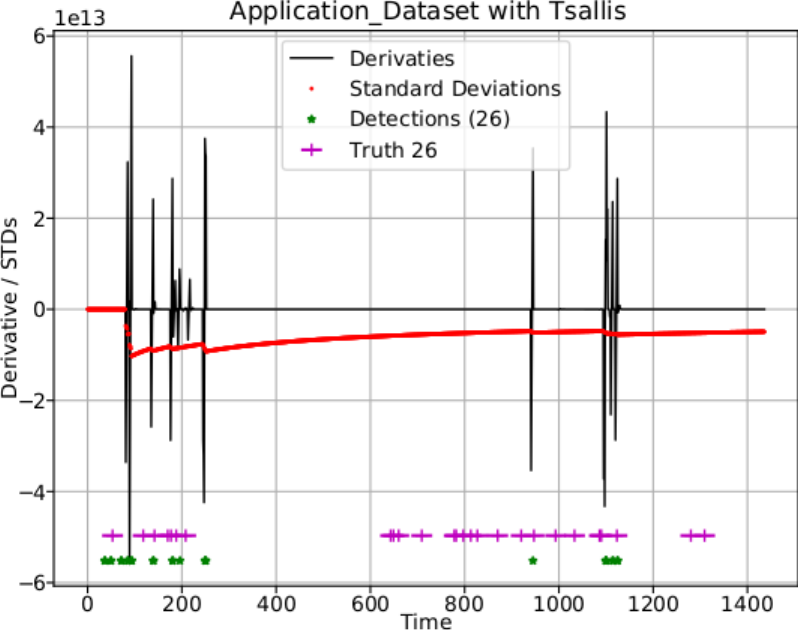}
\includegraphics[scale=0.32]{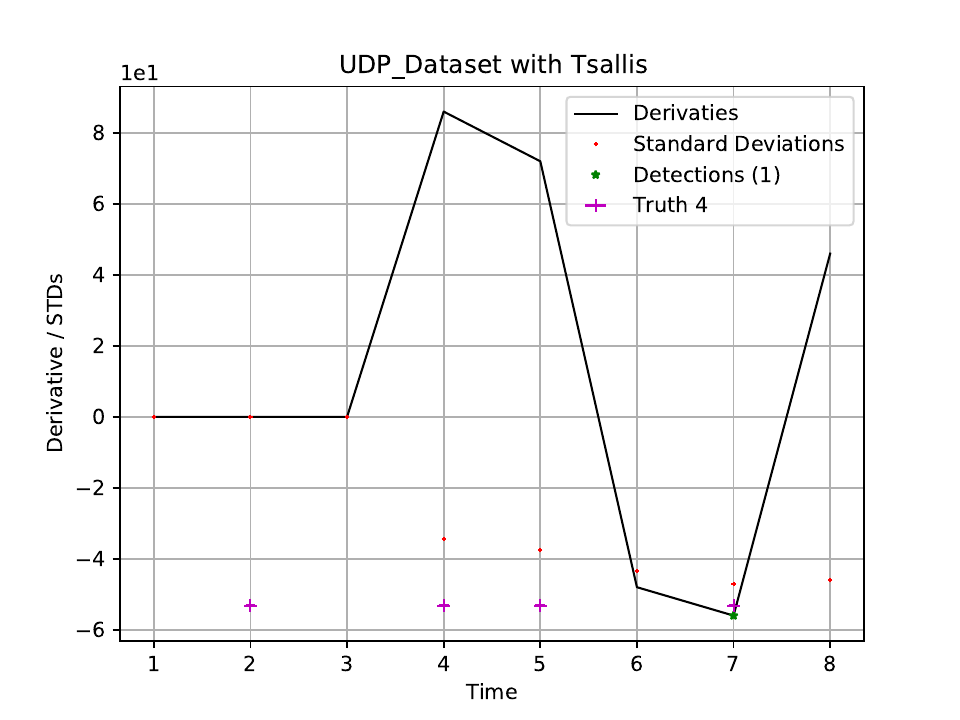}
\includegraphics[scale=0.32]{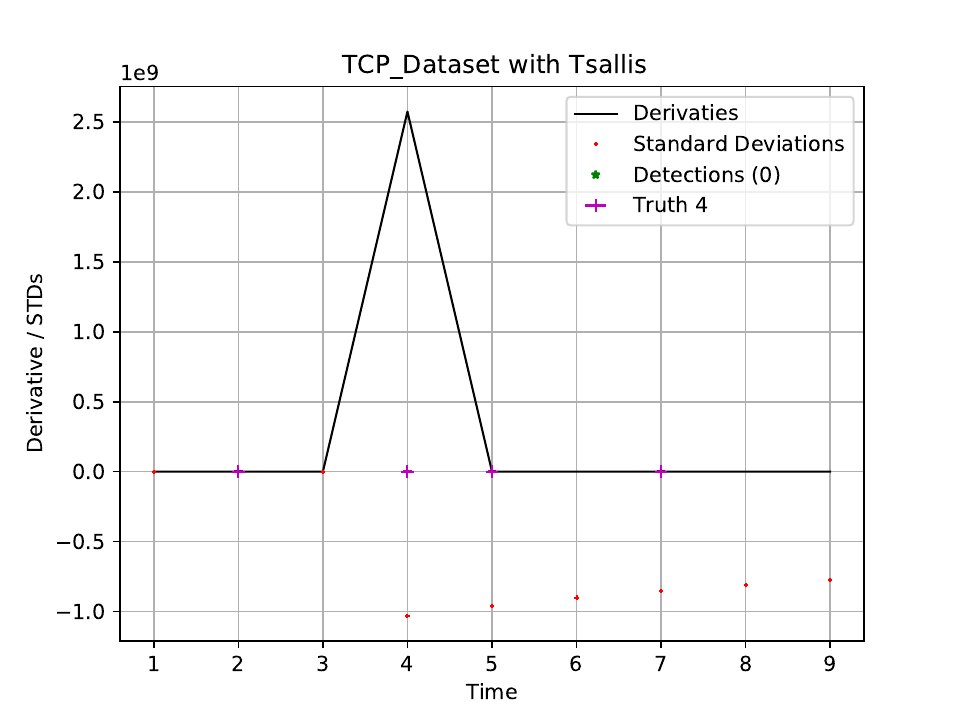}
\includegraphics[scale=0.33]{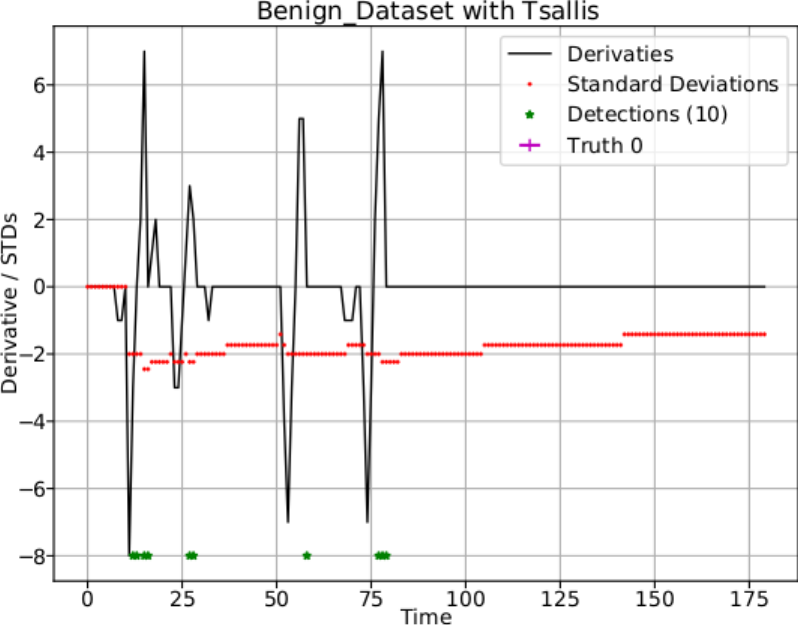}
\includegraphics[scale=0.33]{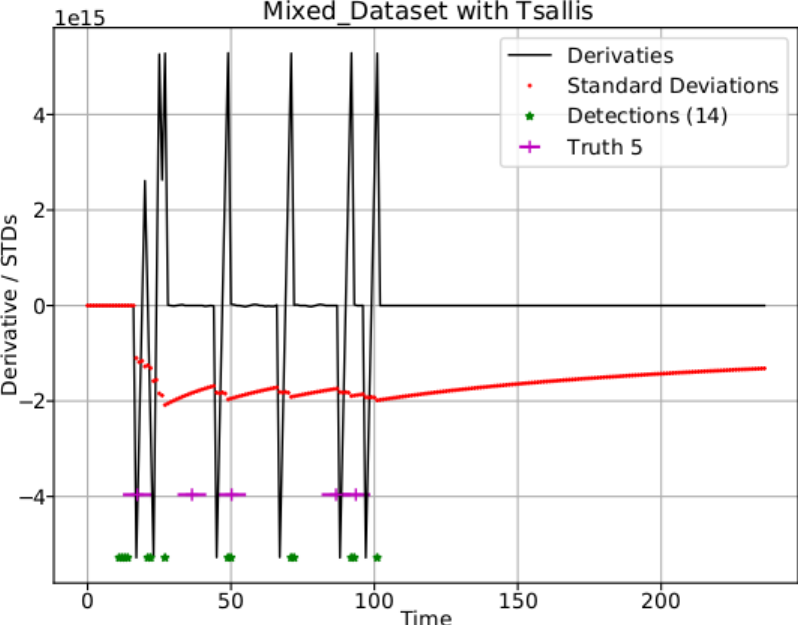}
\caption{The detections and derivatives of DoDGE with 5\% sampling ratio and only integer operations. }
\label{fig:samplederivativeres}
\end{figure*}
\begin{figure*}[ht!]
\centering
\includegraphics[scale=0.47]{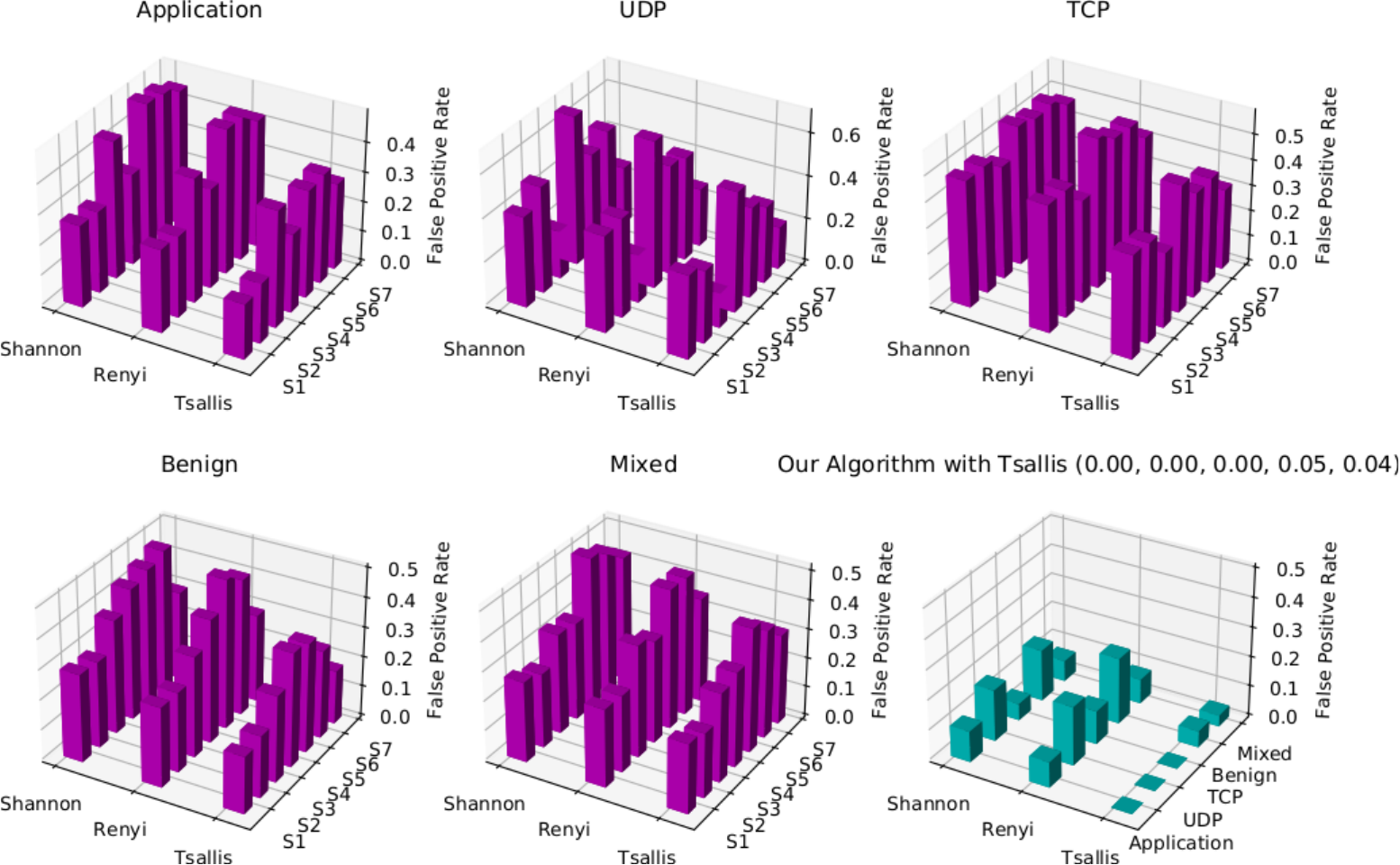}
\caption{Time-based false positive rates. The time-based false positive rates are calculated by dividing the total number of detections minus the number of actual attacks  by the total number of windows. 
DoDGE was run with 5\% sampling ratio and only integer operations.}
\label{fig:falsepos}
\end{figure*}
\subsubsection{{DoDGE Timing and Entropy Analysis}}
Figure \ref{fig:samplederivativeres} shows the derivatives and the standard deviations computed by DoDGE with 5\% sampling ratio and integer operations. 
It also shows the timing of the actual attacks and the detections made by DoDGE. We see that DoDGE produces low number of false positives with the help of the derivatives of the lines of best fit and the standard deviation of the derivatives. 
\begin{table}[ht]
    \centering
    \caption{Detection performances of ML and DoDGE.}
    \begin{tabular}{|l|c|c|c|c|}
    \hline
Algorithm	& Accuracy & Precision & Recall & Balanced Accuracy \\ \hline
SVC &99.2\% &99.2\% & 99.9\% & 50.2\% \\\hline
DT &99.2\% & 99.4\% &99.9\% & 61.6\% \\\hline
RF  & 99.3\% & 99.3\% & 99.9\% & 59.1\% \\\hline
KN & 12.1\% & 97.4\% & 11.8\% & 37.1\% \\\hline
GB & 99.2\% & 99.4\% & 99.8\% & 61.2\% \\\hline
LR & 99.2\% & 99.2\% & 100\% & 50.1\% \\\hline
CONV  & 99.2\% & 99.2\% & 100\% & 50.0\% \\\hline
LSTM & 99.2\% & 99.2\% & 100\% & 50.0\% \\\hline
GRU & 99.2\% & 99.2\% & 100\% & 50.0\%  \\\hline
ED & 99.2\% & 99.2\% & 99.9\% & 49.9\%  \\\hline
\textbf{DoDGE} & \textbf{75.7\%} & \textbf{100\%} & \textbf{75.5\%} & \textbf{99.3\%}  \\\hline
    \end{tabular}
    \label{tab:mlres}
\end{table}

\subsubsection{{ML Comparison}}
We compare DoDGE to ten different ML models that we train and test on the two-day dataset \cite{mlcompdataset}. We train a linear support vector machines (SVC), a Decision Trees (DT), a
Random Forest (RF), a K-Neighbors (KN), a Gradient Boosting (GB), a Logistic Regression (LR),
a Convolutional Network (CONV), a Long Short-Term Memory (LSTM), a Gated Recurrent Unit (GRU), and an Encoder-Decoder (ED) binary classifier on the data from the training day. We then tested them on the test day data. We also test DoDGE on this test-day data. The traditional classifiers are built with Scikit-learn Python library \cite{scikit-learn} while the deep neural network classifiers are built with Tensorflow Keras \cite{chollet2015keras}. Traditional classifiers are built with default parameters except the maximum depth for DT and RF is set to 5 and the number of neighbors is set 3.
CONV is built by three 1D convolutional layers, each followed by a maxpool layer. The number of nodes of the layers are 24, 12, and 6 respectively. The activation function for these layers are RELU. LSTM and GRU are built with 32 nodes with built-in Keras models. LSTM and GRU include a recurrent dropout of 0.25 and are followed with a dropout layer with 0.5. Dropouts are used to prevent over-fitting. The Encoder-Decoder (ED) model is built with three Dense layers with 100, 30, and 100 nodes with RELU activation.
Finally, the last layer for all deep neural networks is a Dense layer with Sigmoid activation function. The loss is set as binary cross entropy, the optimizer is "rmsprop" and the number of epochs is set 5. 

Table \ref{tab:mlres} shows the results for the comparison. We set all ML models except KN achieve an accuracy, precision and recall of 99\% to 100\%. However, since the dataset is highly unbalanced, these typical metrics are not suitable and are misleading. Among 4.3 million instances only 35772 instances are benign (0.8\%). Therefore, in this case, the right metric is to used \textbf{balanced accuracy} \cite{BalancedAcc, KelleherMacNameeDArcy15}. It is defined as $\frac{1}{2} \times (\frac{TP}{TP+FN} + \frac{TN}{TN+FP})$ where TP, FP, TN, FN is true positives, false positives, true negatives, and false negatives, respectively. It is no surprise that almost all classifiers achieve close to 100\% accuracy where the number of benign instances is very small relative to the number of total instances. We see that all ML models have a balanced accuracy less than 62\% while DoDGE has a
balanced accuracy of 99\%. This means that DoDGE outperforms ML models in terms of accuracy. 

\subsubsection{{Comparison to Threshold-based Approaches}}
Figure \ref{fig:falsepos} shows the time-based false positive rates for DoDGE and the threshold-based approaches. The time-based false positive rates are calculated by dividing the total number of detections minus the number of actual attacks by the total number of windows. This rate quantifies the fraction of the false detections made by an approach over time. We see that these rates for the threshold-based approaches range from 10\% to 60\%. This means that these approaches make a false detection at least every 10 windows up to more frequently than every other window. 
 On the other hand, DoDGE with Tsallis entropy has rates that are from 0\% to 5\%. 
 The arithmetic mean is 3\%. These results clearly indicate the success and usability of DoDGE.
We see that due to the challenges inherited by the threshold based approaches, DoDGE is superior with our experimental workflow.

\subsubsection{{Flash Events}}
\begin{figure*}[ht!]
\centering
\includegraphics[scale=0.21]{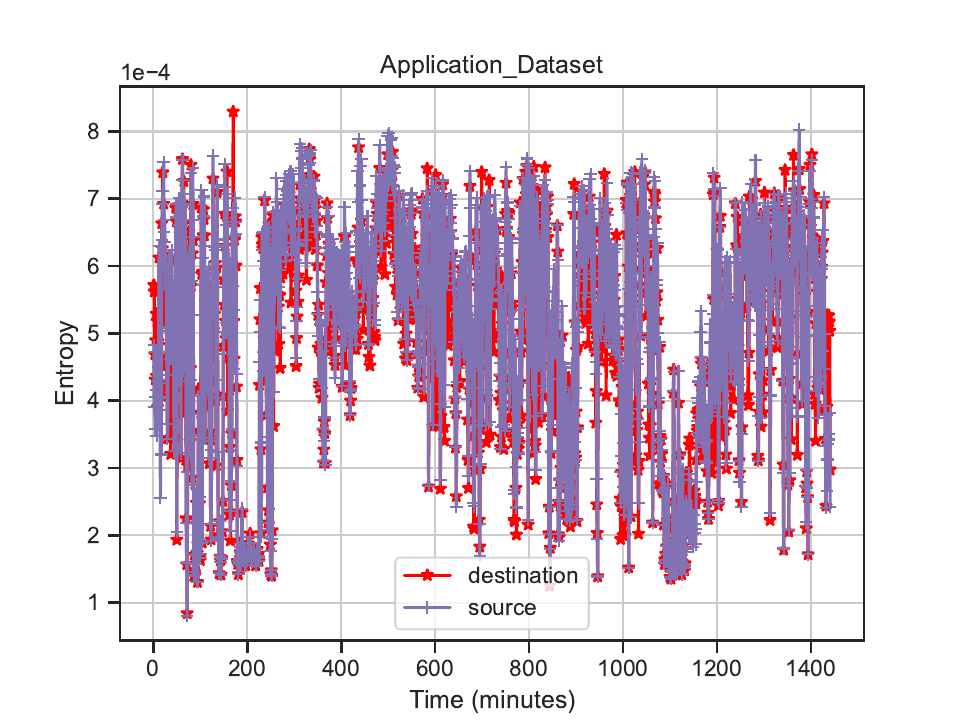}
\includegraphics[scale=0.21]{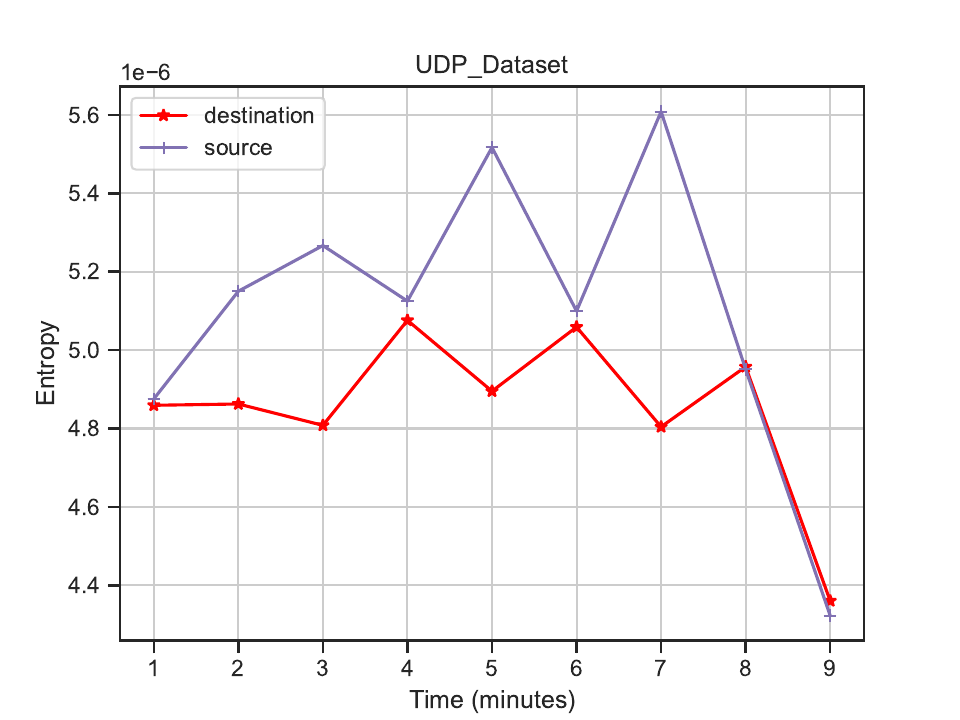}
\includegraphics[scale=0.21]{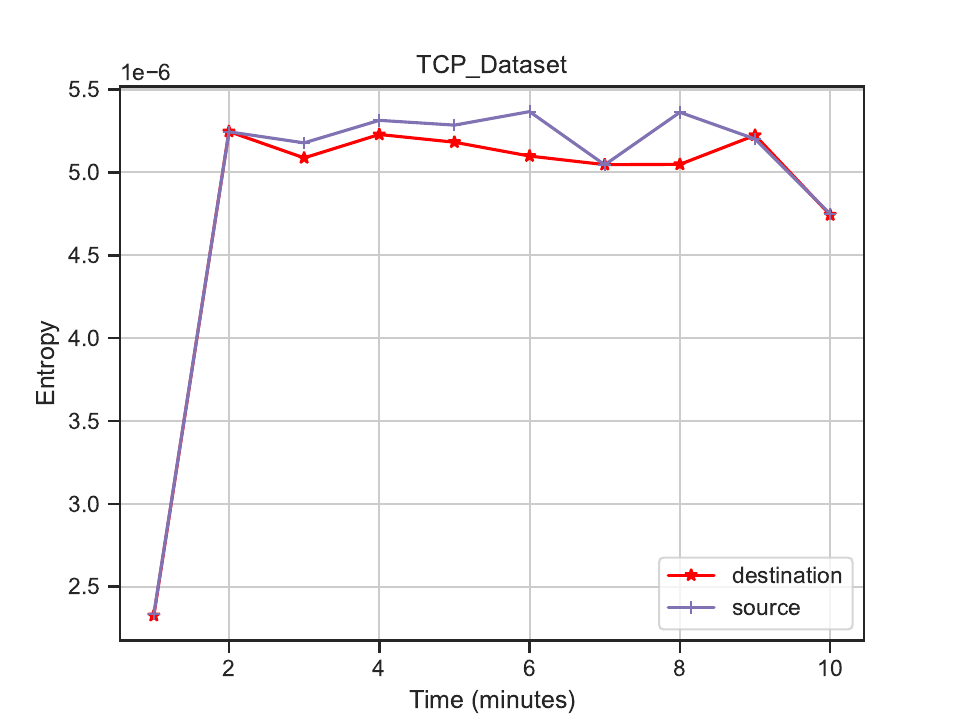}
\includegraphics[scale=0.21]{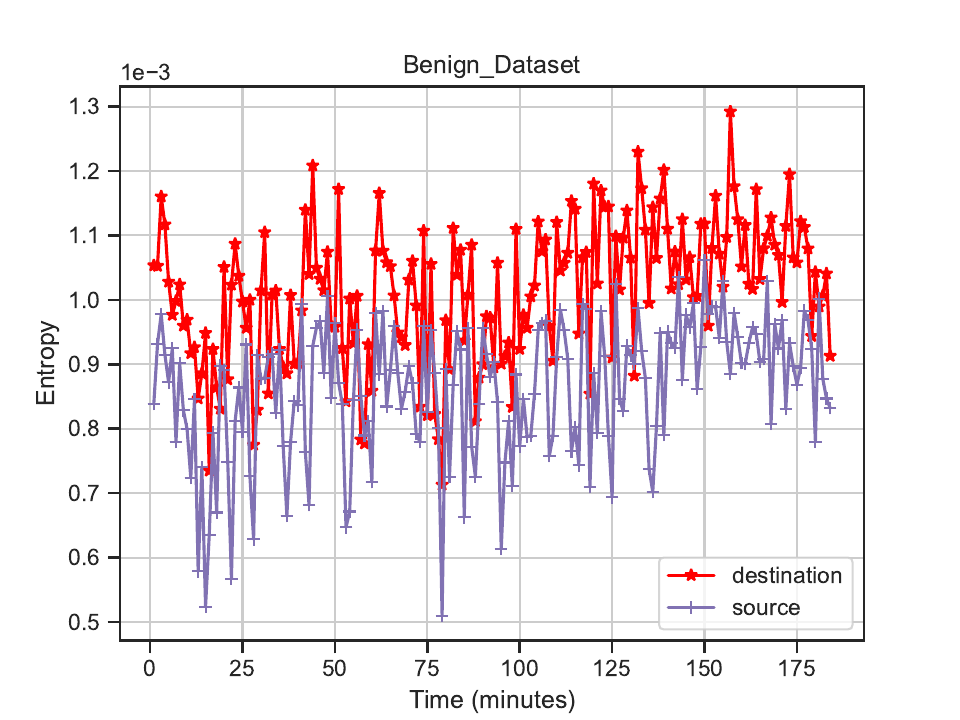}
\includegraphics[scale=0.21]{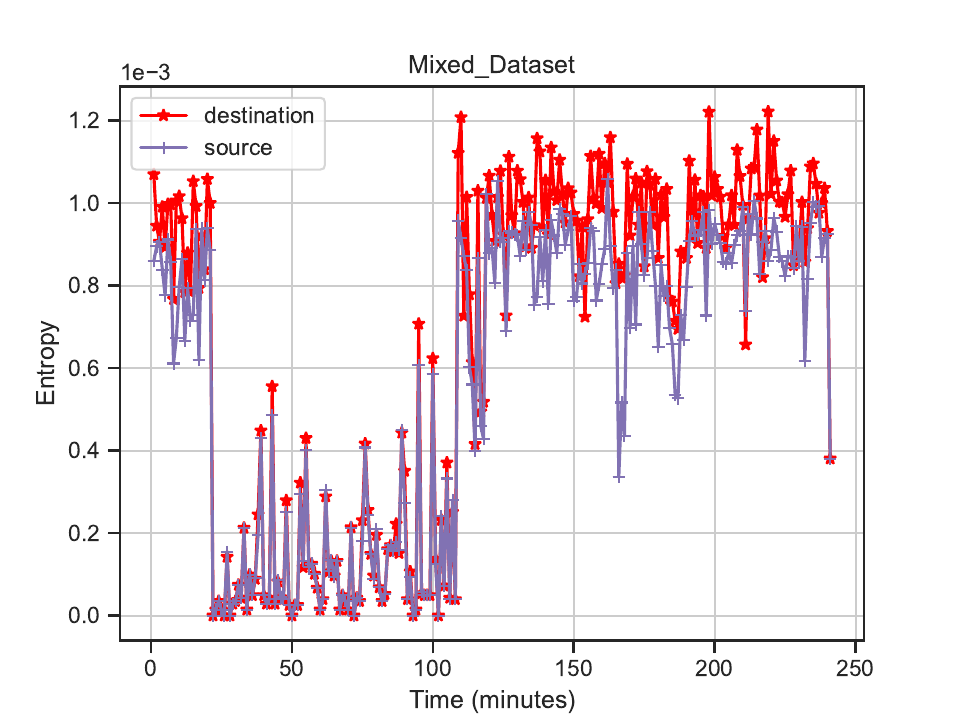}
\includegraphics[scale=0.21]{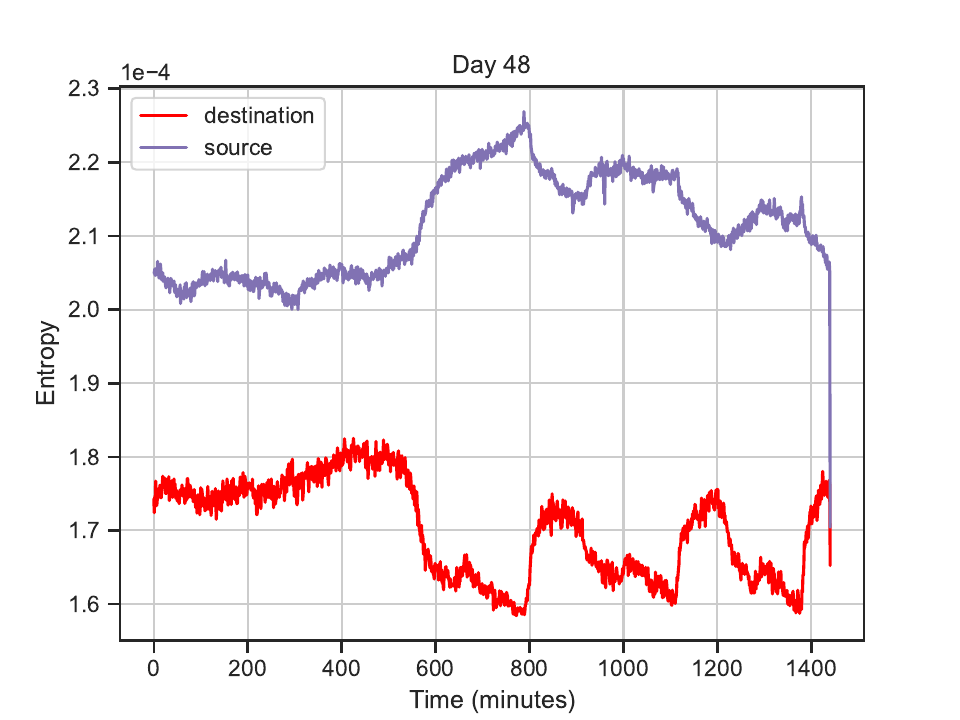}
\includegraphics[scale=0.21]{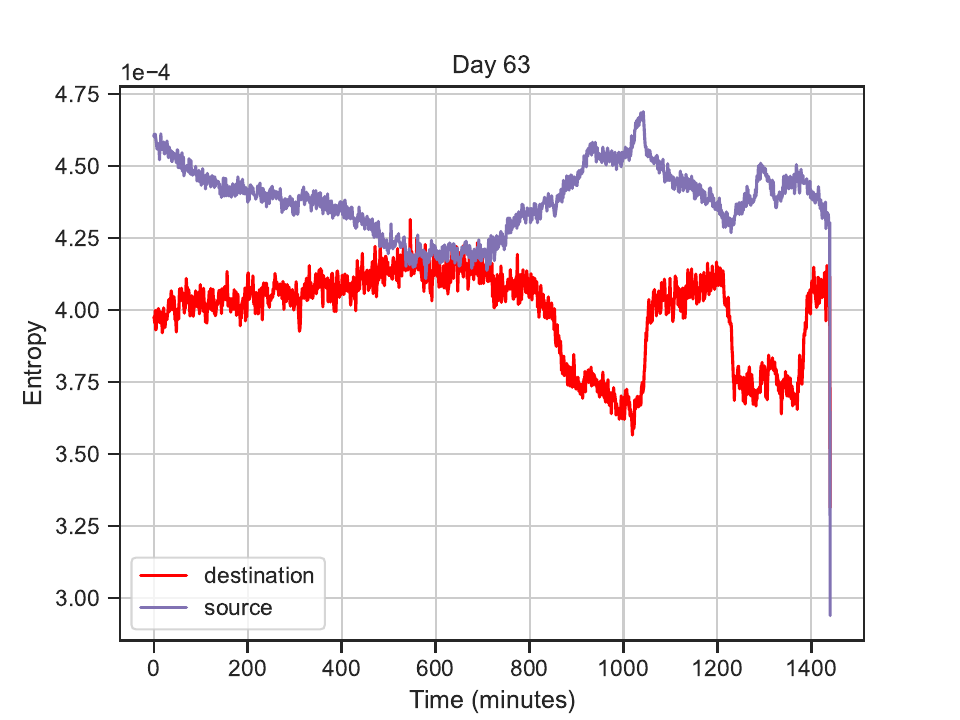}
\includegraphics[scale=0.21]{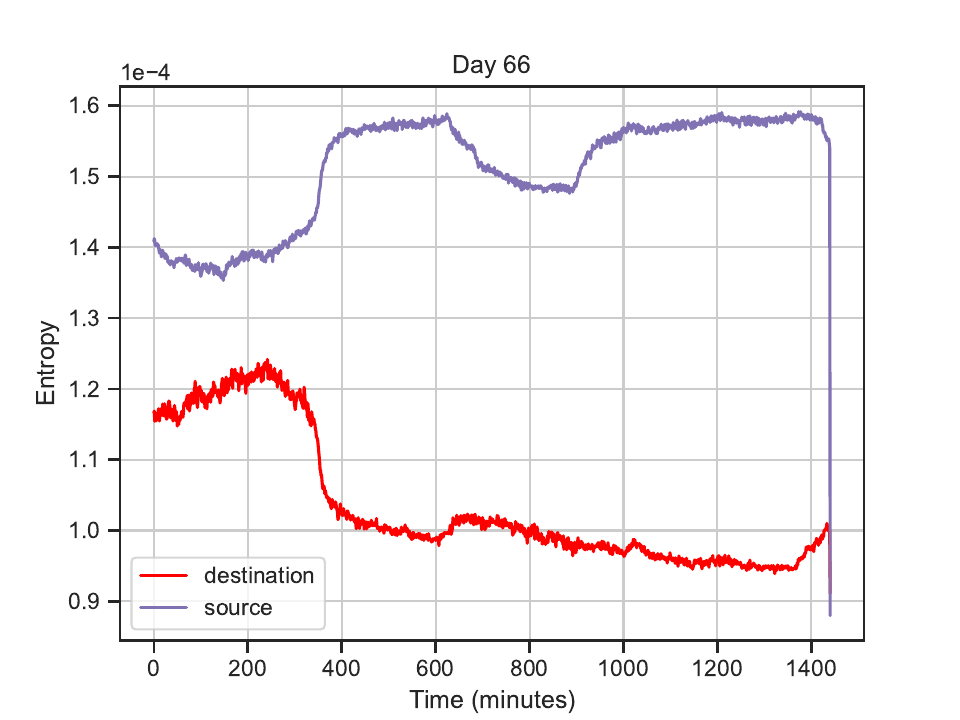}
\includegraphics[scale=0.21]{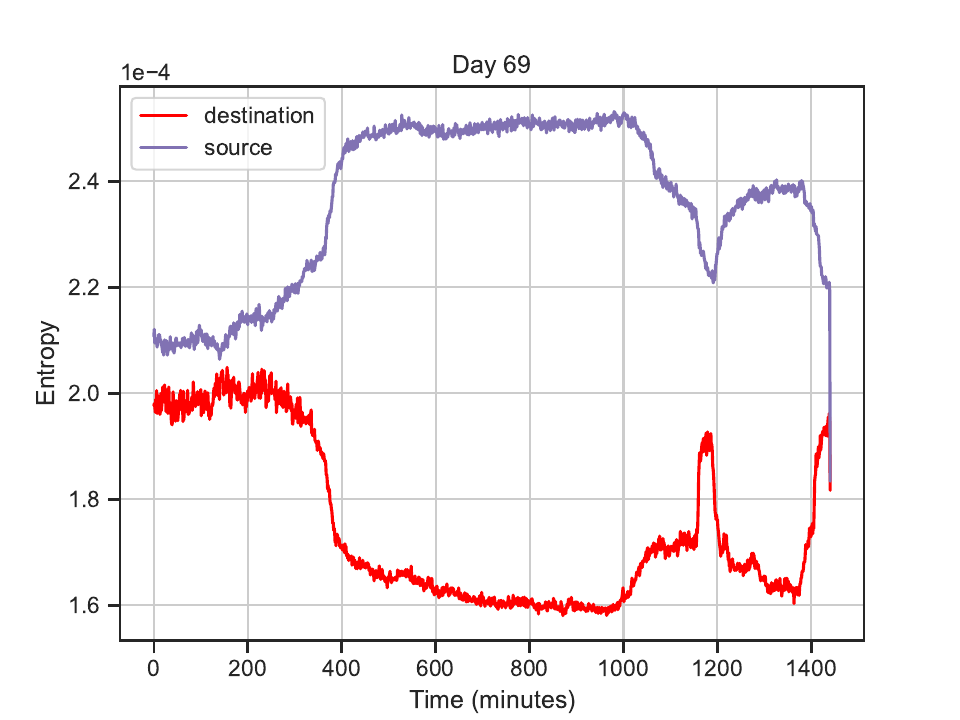}
\includegraphics[scale=0.21]{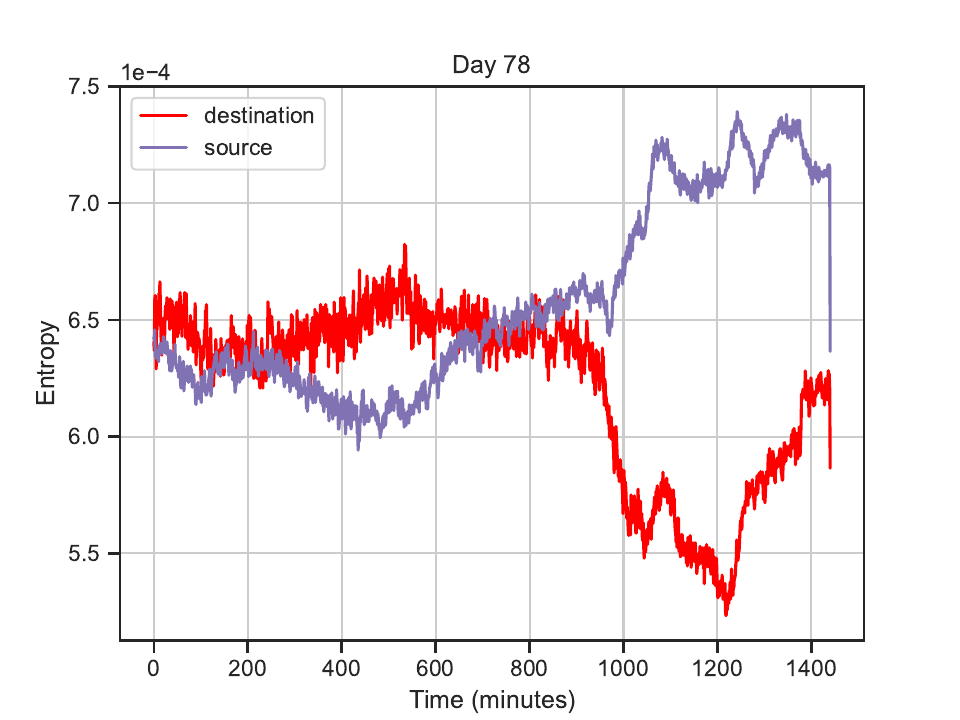}
\caption{Comparison of entropies of DoS Attacks and Flash Events. }
\label{fig:dosflash}
\end{figure*}

Figure \ref{fig:dosflash} 
demonstrates Shannon entropy for the first five datasets which consist of DoS attack traffic and benign Internet traffic,  
and the second five datasets which consist of flash event traffic. We see that while DoS attack traffic does not have a clear difference in source and destination addresses' entropy, the flash events do have source entropy bigger than destination counterparts. Moreover, we see that when flash events occur, i.e, the matches, destination entropy decreases whereas source entropy symmetrically increases. Therefore, these results validate our postulates which guided us to design DoDGE.

During France 98, the matches started at 14:30pm, 17:30pm and 21:00pm for the days with three matches (Day 48 as our dataset), 16:30pm and 21:00pm for the days with two matches (Days 63, 66 and 69) and the final started at 21:00pm (Day 78). Given that unit window is 1 minute, 14:30pm, 16:30pm, 17:30pm and 21:00pm would correspond the windows 14.5$\times$60 = 870, 16.5$\times$60=990, 17.5$\times$60=1150 and 21$\times$60=1260 respectively.
On the bottom row of Figure \ref{fig:dosflash}, on Day 48, the destination entropy decreases around windows 600-800, 900-1100 and 1200-1380 in a clear way, which roughly corresponds to pre-match time and first part of the matches. Symmetrically and closely, source entropy increases during those windows.
On Day 63, there were two matches and we see that destination entropy decreases significantly from about windows 850s and then around windows 1200s. And again source entropy increases closely around the same windows.
On Days 66 and 69, destination and source entropy behave similarly as the two matches start and flash crowds continue toward the end of the day. On the Final Day 78, we see a more pronounced behaviour toward the start of the final. 

Finally, we populate the false positive rates for all ten datasets in Figure \ref{fig:dodgeex}.
We see that the highest false positive rate is from the Final Day 78 which is less then and close to 7\%.
For multiple datasets, i.e. Application, UDP, TCP and Day 48, the rate is 0\%.
The arithmetic mean of DoDGE's false positive rates across all datasets is 1.96\% - less then 2\%.

\subsubsection{{DoDGE Performance}}
\begin{figure}[ht!]
\centering
\includegraphics[width=0.46\columnwidth]{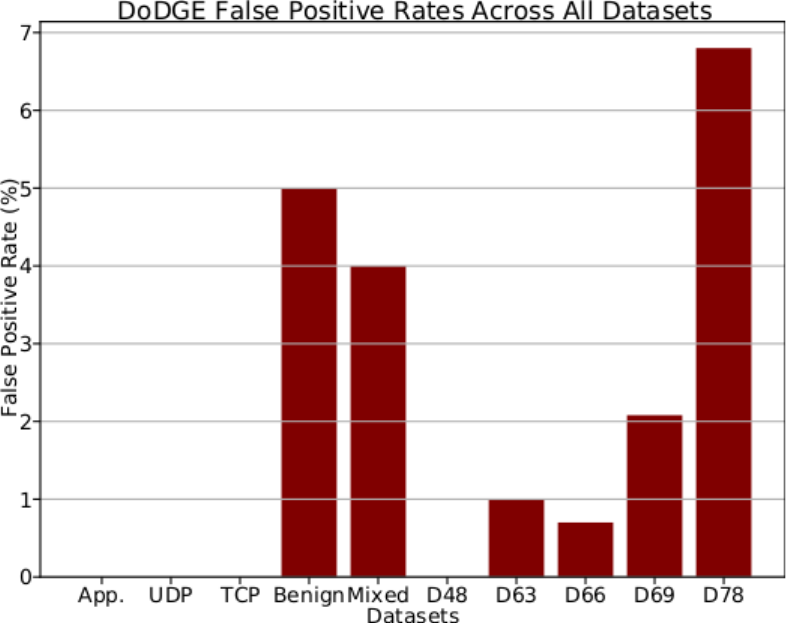}
\includegraphics[width=0.52\columnwidth]{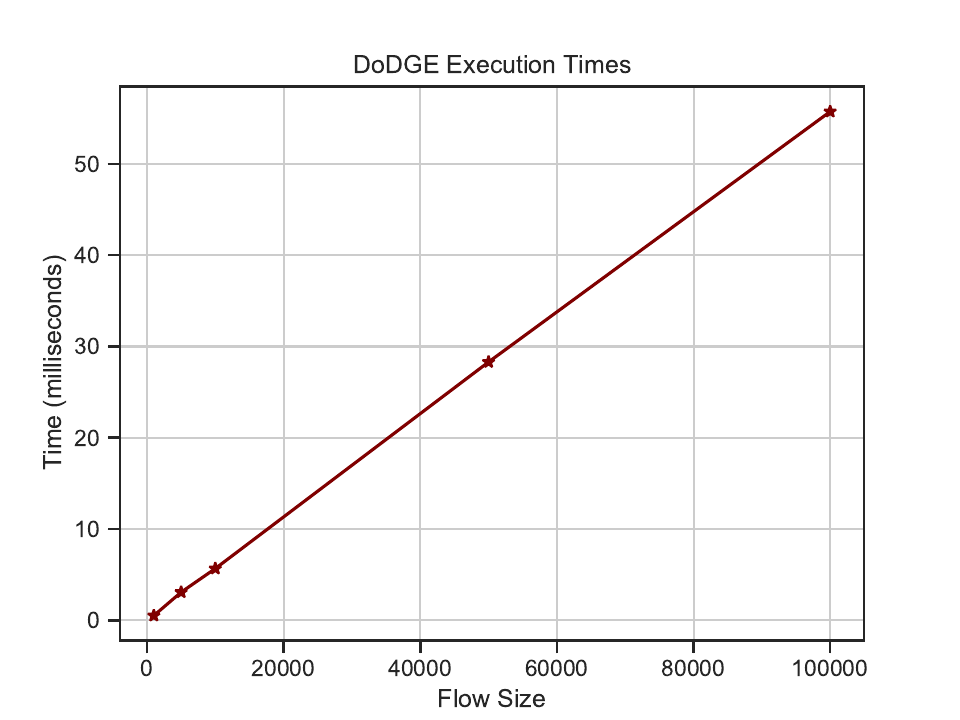}
\caption{DoDGE false positive rates for all datasets (Left). DoDGE execution times with respect to different flow sizes (Right).}
\label{fig:dodgeex}
\end{figure}

Figure \ref{fig:dodgeex} shows the execution times in milliseconds 
of DoDGE with respect to different flow sizes - that
is, the number of address entries during one unit window (average over 100 repetitions).
We see that the computational cost of DoDGE is linear with flow size.
In our prototype C++ implementation, given a vector of source and a vector of destination 
addresses, a DoDGE execution consists of the frequency counting of source and destination addresses 
by using C++ integer maps, getting the values vectors of these two maps, calculating Tsallis entropy for
the two values vectors, performing two lines-of-best-fit to compute the source and destination 
derivatives, applying Wellford's method for dynamic computation of the standard deviation of 
the destination derivatives, and finally the condition checking to differentiate
among DoS attacks, flash events, and benign traffic.
In terms of memory consumption, our implementation uses four vectors whose size is the flow size, and uses two integer to integer maps for the
frequencies of the flow entries.

\section{Related Work}
\label{related}
In this work, we focus on information-theoretical approaches. \cite{8778219} and \cite{JOLDZIC201627} propose an entropy-based approach to detect DoS attacks. 
The entropy used in these studies is Shannon entropy.
Similar to our method, the entropy is defined based on destination address. We differently use generalized entropy measures with differential analysis to improve the detection performance while removing the usage of thresholds.

In terms of the usage of generalized entropies in the existing literature,
Behal and Kumar \cite{BEHAL201796} introduce $\phi$-entropy and $\phi$-divergence metrics
to detect and distinguish various types of DoS attacks and flash events. They report the $\phi$ metrics perform better than traditional generalized entropy and divergence metrics. Moreover, $\phi$-divergence exceeds $\phi$-entropy in detection accuracy. Basicevic et. al. \cite{Basicevic} study various DoS attack detectors based on different entropy types using source side features and report that detectors based on Tsallis \cite{tsallis1988possible} - in agreement with our study - and Bhatia-Singh \cite{Bhatia-fdiv} entropies perform best. In comparison, we use differential and statistical analysis to signal an attack which is hyper-parameter free and thus is robust. Moreover, we study many real-world DoS attack types whereas they study only SYN floods.

Some research studies use other aspects of entropy such as joint and conditional entropy. Mao et. al. \cite{joint-entropy} propose a joint-entropy based detector for DoS flood attacks. While they report that joint-entropy generally leads to better performance compared to single entropy, the overall algorithm requires more memory and although not asymptotically, it is also slower than single entropy calculations. One major drawback of their approach is the threshold is static. 
\cite{conditional} proposes using conditional entropy to detect DoS attacks. However, conditional entropy is not scalable because it has quadratic time and memory complexity. 

Other than entropy, some information-theoretical approaches make use of different metrics. 
Behal et. al. \cite{BEHAL2021291} studies $\phi$-divergence and develops an distributed method. This mechanism is distributed over the Point of Presence of ISPs. A centralized server aggregates the entropies and decides if there is an DoS attack or a flash event. In case of DoS attacks, they are further classified as low-rate or high-rate. The main drawback of this approach is that the centralized server becomes single-point-of-failure. 
As with the existing research, the success of their method depends on the optimization of window size and thresholds.
\cite{JAZI201725} presents a cumulative sum metric for high and low volume DoS attacks while also studying the impact of various types of sampling on the detection performance. Their method requires the optimization of window size and thresholds which inhibits the performance of their method in general settings. 

\section{Conclusion and Future Work}
\label{conclusion}
In this work, we propose a novel method, named DoDGE, based on a generalized entropy measure, Tsallis entropy,  to detect DoS attacks and differentiate them from flash events. 
DoDGE progressively uses the derivatives of the entropy progressions to detect attacks, flash or benign events. The differentiation between DoS attacks and flash events is based on the symmetry between 
the entropies of the source and destination addresses. Our postulates on which DoDGE were
developed are i) the entropy of the destination progression decreases when a DoS attack happens, ii) this decrease also holds for flash events, and iii) a necessary condition for differentiating attacks from flash events is the existence of the symmetrical behavior of
the entropy of the source progression, which is its increase.
DoDGE is by construction is embarrassingly distributed: it is to be deployed to 5G edge nodes or Internet routers. In our thorough evaluation, we see it achieves false positive rates that are less than 7\% in all cases and on average 1.96\% across ten real-world datasets. It outperforms
threshold-based methods by two orders of magnitude in terms of false positives and outperforms ML models significantly in terms of balanced accuracy - 99\% to the highest score of 62\%.

\section*{Acknowledgement}
This work was supported by the U.S. DOE Office of Science, Office of Advanced Scientific Computing Research, under award 66150: "CENATE - Center for Advanced Architecture Evaluation" project. The Pacific Northwest National Laboratory is operated by Battelle for the U.S. Department of Energy under contract DE-AC05-76RL01830.

\bibliographystyle{IEEEtran}
\bibliography{references}

\end{document}